\documentclass[doublecol]{epl2} 
\usepackage{epsfig}
\usepackage{epstopdf}
\usepackage{ifpdf}
\usepackage{graphics}
\usepackage{amssymb,amsmath}
\usepackage{subfigure}
\usepackage{mathrsfs}
\usepackage{bbm}

\title{Relativistic linear stability equations for the nonlinear Dirac equation in Bose-Einstein condensates}
\shorttitle{Relativistic linear stability equations for the nonlinear Dirac equation}

\author{L. H. Haddad and L. D. Carr}

\institute{  Department of Physics, Colorado School of Mines, Golden, CO
80401, USA\\
}

\abstract{We present relativistic linear stability equations (RLSE) for quasi-relativistic cold atoms in a honeycomb optical lattice. These equations are derived from first principles and provide a method for computing stabilities of arbitrary localized solutions of the nonlinear Dirac equation (NLDE), a relativistic generalization of the nonlinear Schr\"odinger equation. We present a variety of such localized solutions:  skyrmions, solitons, vortices, and half-quantum vortices, and study their stabilities via the RLSE. When applied to a uniform background, our calculations reveal an experimentally observable effect in the form of Cherenkov radiation. Remarkably, the Berry phase from the bipartite structure of the honeycomb lattice induces a boson-fermion transmutation in the quasi-particle operator statistics.}

\vspace{-.5pc}
\begin{document}

\maketitle
Progress in condensed matter and particle physics has been periodically marked by significant mutual exchanges between the two disciplines, many proposals for which are realized in model systems of ultracold quantum gases in optical lattices~\cite{garay2000,rapp2007,yuYue2008}.~Recent active areas of research include holographic dualities such as AdS/CFT~\cite{hartnoll2010},~theoretical constructions of superstrings in ultracold quantum gases~\cite{snoek2009}, chiral confinement in quasi-relativistic Bose-Einstein condensates (BECs)~\cite{merkl2009}, and our own derivation of the \emph{nonlinear Dirac equation} (NLDE) describing ultracold bosons in a honeycomb optical lattice~\cite{carr2009g}.~Our investigation into relativistic effects in BECs is motivated by this spirit of cross fertilization with the aim of tying in theory to experiment.

In this Letter, we develop the \textit{relativistic linear stability equations} (RLSE) for the NLDE. Moreover, we find emergent nonlinear localized solutions~\cite{kevrekidisPG2008} to the NLDE, including solitons, vortices, skyrmions, and half-quantum vortices, the latter so-far unobserved in BECs.~Although most of these objects have been studied in multicomponent BECs, such models lie within the usual Schr\"odinger many-body paradigm.~In contrast to this paradigm, our investigations reside within a relativistic framework in which the elementary excitations are governed by a Dirac-like equation. This provides a fundamentally different context distinguished by the presence of a Berry phase, so that exchange of two vortices leads to integer or half-integer exchange statistics. It is not surprising that the elementary excitations in our theory exhibit a rich structure: a Dirac-like dispersion which obeys either bosonic or fermionic statistics depending on the strength of the contact interaction.~Consequently, in order to determine the quasi-particle states and energies we cannot rely on the Bogoliubov-de Gennes equations (BdGE) since these are based on nonrelativistic quantum mechanics.~Instead, we derive, from first principles, the RLSE which give the correct low energy dynamics for an arbitrary background condensate. The RLSE are reducible to the BdGE in certain limits, and so may naturally be considered relativistic generalizations of the latter.  Based on the RLSE we predict Cherenkov radiation that can be measured in experiments: the combination of lattice and particle interactions results in a rich spatial distribution that is not seen in the BdGE for the uniform case~\cite{fetterAL1972}.

\begin{figure}
\vspace{0pc}
\begin{tabular}{c c c}
 \hspace{-1.2pc}    \hspace{-.2pc} \vspace{-.4pc} &  \hspace{-1.2pc}  \includegraphics[width=4cm]{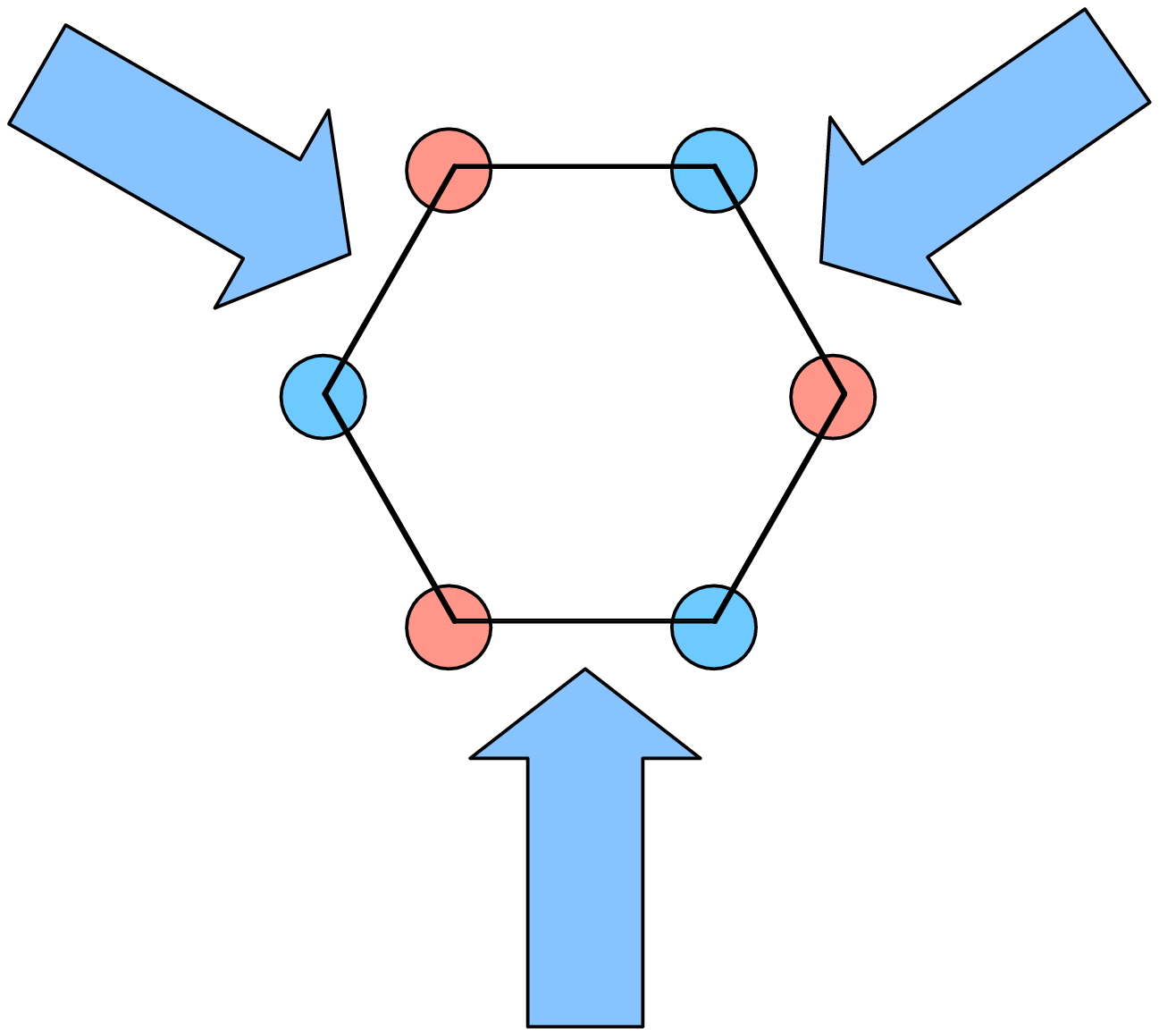} \vspace{-2.4pc}  & \hspace{-3.5pc} \vspace{-1pc}  \tabularnewline
 \hspace{-4pc} \vspace{.41pc} \small{E(k)} & \hspace{-5.7pc}  \vspace{-1pc}(b)  & \hspace{-3pc} \vspace{.4pc} $\small{\textrm{E}(\textrm{k}_\textrm{x},\textrm{k}_\textrm{y})}$ \tabularnewline 
  \hspace{1pc} \includegraphics[width=3cm]{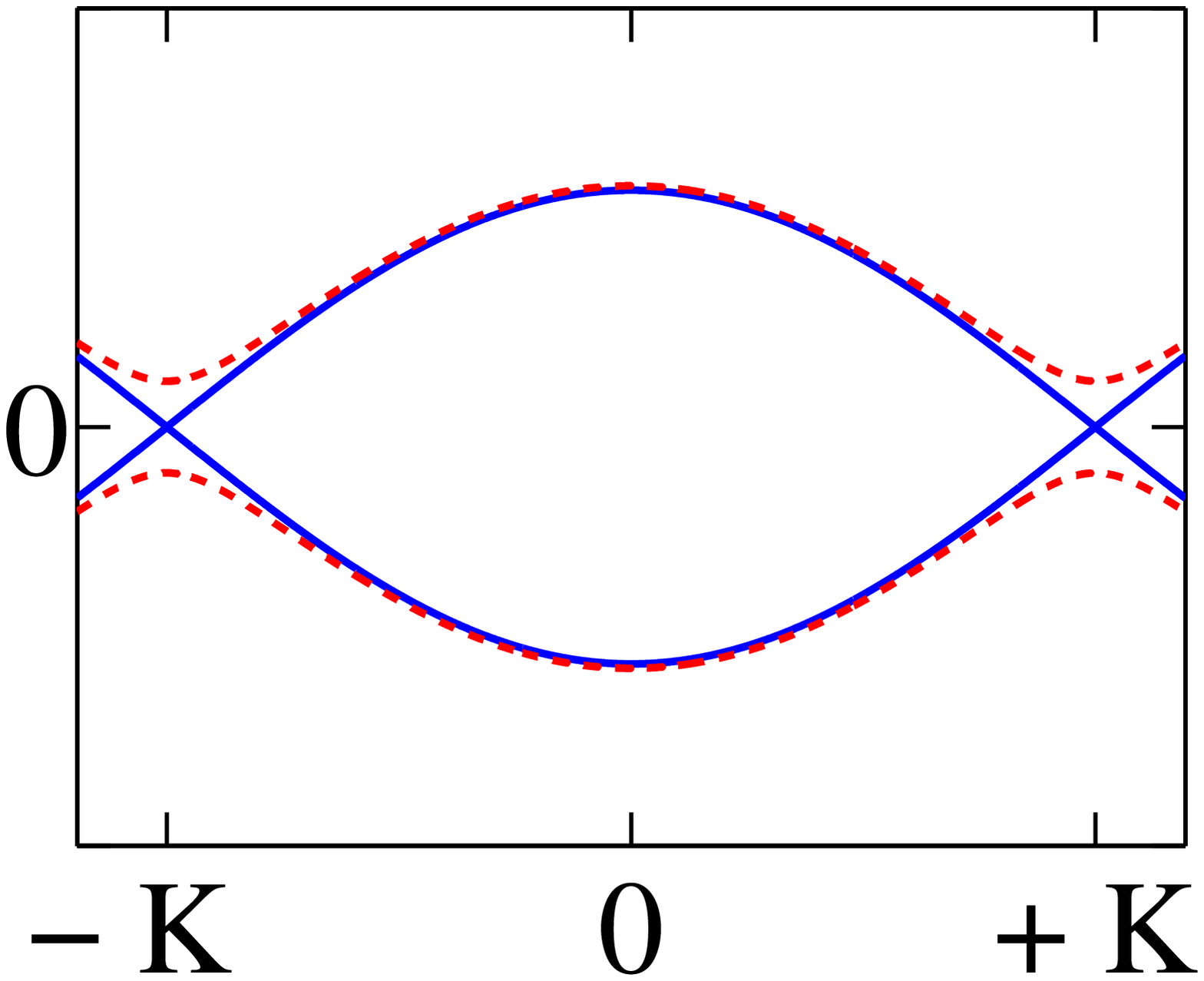} \hspace{-2pc} \vspace{0pc} &  \hspace{-.7pc}        \vspace{0pc}  & \vspace{-5.4pc}     \tabularnewline
  \hspace{-.7pc}        \vspace{-.1pc}  &    \hspace{-.7pc}        \vspace{-.1pc}  &  \hspace{-5pc}\includegraphics[width=2.7cm]{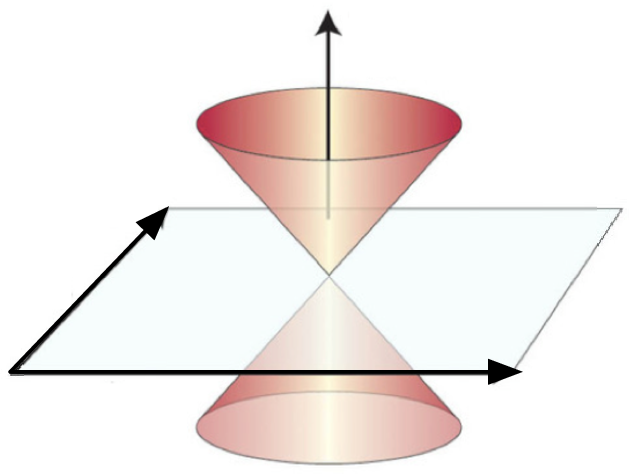} 
\hspace{-1.8pc}  \vspace{-.7pc} \hspace{-1pc}      \tabularnewline 
\hspace{1pc} (a)  & \hspace{6pc} (c)  &  \tabularnewline 
\end{tabular}
\caption{ \emph{The honeycomb optical lattice}.~(a) Cross section of the band structure showing $K$ and $K'$ points for gapped and ungapped systems.~(b) The velocity and acceleration of the lattice, with A and B sub-lattices, are functions of the frequency offsets for interfering lasers.~(c) Two-dimensional Dirac cone at $K$ and $K'$.}  
\label{fig.1}
\vspace{-1.1pc}
\end{figure}

In the laboratory, the NLDE can be obtained by cooling bosons into the lowest Bloch band of a honeycomb optical lattice~\cite{soltan-panahi2010}; the lattice is constructed by establishing three phase-locked interfering laser beams in a plane while freezing out excitations in the vertical direction as in Fig.~\ref{fig.1}. To obtain the desired Dirac structure, particles are first condensed into the lowest energy state (zero crystal momentum) of the lattice and then adiabatically translated to the Dirac point at the band edge  (see Fig.~\ref{fig.1}) by adiabatically tuning the relative phases between the laser beams.  We emphasize that the Dirac point, which is key to the NLDE and our predictions, is maintained even in the presence of the shallow harmonic trap endemic to atomic BECs~\cite{kusk2010}.~Notably, the NLDE may also be obtained by the alternative method of using a square optical lattice with a staggered gauge field induced by a time-dependent optical potential~\cite{Lim2009}.~A similar arrangement using only bosons is expected to show the same low energy structure as in our model. 
Nonlinear phenomena in BECs have been studied extensively over the past decade~\cite{kevrekidisPG2008}, from single-component vortices in rotating, trapped BECs~\cite{feder2000} to complex multi-component order parameters~\cite{muellerEJ2004,kasamatsu2005} resulting from interactions between the different components and the possibility of nontrivial topological windings of the internal symmetry space around a singular vortex core.~Some form of BdGE analysis plays a central role in such constructions as a means of probing stability as well as for gaining a deeper understanding of the low-energy fluctuations. 

\section{Derivation of the RLSE} Since the system we describe in this letter is a BEC confined strictly to two spatial dimensions, it is appropriate to recall the justification for such a construction before presenting the RLSE. For uniform 2D systems the Mermin-Wagner theorem forbids the formation of a true condensate defined by an infinite phase coherence length.~This comes from the fact that the density of states diverges in the 2D case for finite $T$. Instead, one sees the formation of a \emph{quasi-condensate} characterized by local phase coherence restricted to finite size regions. The size of these regions greatly exceeds the healing length so that all of our solutions are realizable in this picture.~However, the inclusion of a harmonic confining potential allows the formation of a true 2D condensate. The potential places a lower bound on the energy for fluctuations and, since it is these long wavelength fluctuations that are responsible for destroying long range order, the trap provides a means of expanding the spatial range of~validity of the mean field description. 
\vspace{-.22pc}

To obtain the low energy excitations of solutions of the NLDE,~we must find the correct set of equations that describe quasi-particle states analogous to the BdGE equations for the general case.~These are obtained from the Hamiltonian for a weakly interacting Bose gas, $\hat{H}= \int \! d{\bf r}\,  \hat{\psi}^\dagger H_0 \hat{\psi} + \frac{U}{2}  \int \! d{ \bf r} \, \hat{\psi}^\dagger\hat{\psi}^\dagger  \hat{\psi} \hat{\psi}$, $H_0 \equiv \hbar^2\nabla^2/2M + V({\bf r})$, and working through four steps~\cite{pitaevskii1961}.~(1) Take $\hat{\psi}=\Psi_c({\bf r})+\delta\hat{\psi}_{\bf q}({\bf r})$ (condensate + quasi-particles), with $\delta\hat{\psi}_{\bf q}$ small.~(2) Impose a constraint on $\Psi_c$ to eliminate linear terms in $\delta\hat{\psi}_{\bf q}({\bf r})$, keep only quadratic terms in $\delta\hat{\psi}_{\bf q}({\bf r})$, and expand as a sum of particle and hole creation operators.~(3) Invoke Bloch-state expansions for $\Psi_c({\bf r})$ and $\delta\hat{\psi}_{\bf q}({\bf r})$ and take the lowest band.~(4) Take the long-wavelength limit while taking momentum with respect to the Dirac point ${\bf K}$, such that $ {\bf k} \ll {\bf q} \ll {\bf K}$, where ${\bf q}$ is the momentum of the condensate relative to the Dirac point ${\bf K}$ and ${\bf k}$ is the quasi-particle momentum measured relative to {\bf q}, and finally diagonalize the quasi-particle part of the Hamiltonian. One finds the RLSE:
\vspace{-.99pc}
\begin{eqnarray}
\hspace{-1pc}&&\tilde{\mathscr{D}}  {\bf u}_{\bf k}  - U \tilde{{\Psi}}   {\bf v}_{\bf k} =  \tilde{ E}_{\bf k} {\bf u}_{\bf k} , \label{eqn:RLSE1} \\
\hspace{0pc}&&\tilde{\mathscr{D}}^*  {\bf v}_{\bf k}  - U \tilde{{ \Psi }}  {\bf u}_{\bf k} =- \tilde{E}_{\bf k} {\bf v}_{\bf k}  \; ,  \label{eqn:RLSE2}\end{eqnarray}
\vspace{-.5pc}
\noindent where the matrix coefficients are defined as
\begin{eqnarray}
\hspace{-3pc}\tilde{\Psi}&\equiv& \text{diag}(    \left| \Psi_A \right|^2 ,   \left| \Psi_B \right|^2  ),  \\
\hspace{-3pc}\tilde{ E}_{\bf k} &\equiv& \text{diag}(   E_{\bf k}   ,   E_{\bf k}   ),    \\
\hspace{-3pc} {[}  \tilde{\mathscr{D} } {]}_{1,1}  &\equiv&   m_{\mathrm{eff}} - \mu+ 2 U \left| \Psi_A \right|^2 - i   \nabla \phi_A \cdot \nabla \nonumber \\
\hspace{-3pc}&&+ \left|\nabla \phi_A \right| - i \left( \nabla^2 \phi_A\right),  \\
\hspace{-3pc} {[}  \tilde{\mathscr{D} } {]}_{2,2}  &\equiv&   m_{\mathrm{eff}} - \mu + 2 U \left| \Psi_B \right|^2 - i   \nabla \phi_B \cdot \nabla \nonumber \\
\hspace{-3pc}&&+ \left| \nabla \phi_B \right| - i \left( \nabla^2 \phi_B\right),  \\
 {[} \tilde{\mathscr{D} }  {]}_{1,2}  &=&  {[} \tilde{\mathscr{D} }  {]}_{2,1}^* \equiv    \mathcal{D}^*.
\end{eqnarray}
\noindent Here, $\mathcal{D}\!=\!(\partial_x+i\partial_y)$ is the single particle Dirac operator.~Also, $\Psi\!=\!(\Psi_A , \Psi_B)$ is the BEC order parameter at the ${\bf K}$ Dirac point, with normalization on sublattice components $\!\int d{\bf r}\, (|\Psi_A|^2+|\Psi_B|^2)=1$.~Analogous equations hold for the inequivalent Dirac point at $-{\bf K}$. Cast in this highly compact form, Eqs.~(\ref{eqn:RLSE1})-(\ref{eqn:RLSE2}) are reminiscent of the BdGE and may be solved for the \emph{spinor quasi-particle amplitudes} ${\bf u}_{\bf k}({\bf r}) =  [  u_{{\bf k},A}({\bf r})   ,  u_{{\bf k},B}({\bf r}) ]^T$ and  ${\bf v}_{\bf k}({\bf r}) = [ v_{{\bf k},A}({\bf r})   ,  v_{{\bf k},B}({\bf r})]^T$ and the quasi-particle energy $E_k$.~The components of these 2-spinors represent quantum fluctuations of the sublattice condensate order parameters $\!\Psi_A\!$ and $\!\Psi_B\!$ which in general are nonuniform $\mathbb{C}$-functions on the plane. The presence of the local phase of the condensate $\phi_{A(B)}({\bf r})$ indicates the complex interaction between the local superfluid velocity of the condensate $ { \bf v}_{s, A(B)}({\bf r}) \equiv  \nabla \phi_{A(B)}({\bf r})$ and the spinor quasi-particles ${\bf u}_{\bf k}({\bf r})$ and ${\bf v}_{\bf k}({\bf r})$. We have taken $\hbar=c_l=1$ for simplicity, where $c_l$ is the effective speed of light in the NLDE.~Note also that we have included an effective mass $m_{\mathrm{eff}}$ (anisotropic lattice) that competes with the chemical potential $\mu$. 

The RLSE exhibit a positive-negative energy inversion symmetry which is found by complex conjugating Eqs.~(\ref{eqn:RLSE1}) and (\ref{eqn:RLSE2}) followed by a spatial parity inversion.~This is similar to the BdGE where negative energy modes are allowed by symmetry but are ignored since the underlying Hamiltonian is positive-definite.~It is important to note that for a moving condensate, the negative-energy modes cannot be removed and are crucial indicators of Cherenkov radiation.~However, in our case, the Dirac Hamiltonian is \emph{not} positive-definite since our theory is defined at zero lattice energy, not the lowest energy Bloch state, so we must respect the presence of both energy raising \emph{and} lowering modes.~Another important feature is that the RLSE are reducible to the BdGE when the local lattice potential energy is the main contributor to the condensate chemical potential and the condensate is slowly varying (quasi-uniform background), i.e., $| \mu| >> U$, $E_j \to | \mu | \approx | \Sigma_0|$, where $\Sigma_0 =\Sigma_{0 A(B)}\equiv - \int \! d{\bf r}\,  w_{A(B)}^* H_0 w_{A(B)}$ is the local self energy for an arbitrary lattice site, with $w_{A(B)}=w({\bf r} - {\bf r}_{A(B)})$ the Wannier functions.
\vspace{-.4pc}

\section{Quasi-particle Operator Statistics} Since we use only bosons in our construction, and we have shown that these collectively produce the spinor structure of Dirac theory, it is natural to ask what are the quantum statistics for collective excitations.~When quantum fluctuations are weak, quasi-particle states are superpositions of direct products of bosonic operators and spinor single-particle states: operators retain their bosonic structure. In contrast, when the depletion of the condensate becomes significant, quasi-particle operators at each lattice site become sums of superpositions of terms each with a different phase, i.e., the single-particle bosonic operators $\hat{b}$ and $\hat{b}^\dagger$ become entangled with the on-site phase.~We can illustrate this more rigorously.~Consider a BEC in the thermodynamic limit: $\!1/N \!\to \! 0$ and $N/\mathcal{M} = \mathrm{constant}$, where $N$ is the total number of particles and $\mathcal{M}$ is the number of lattice sites. Also, we focus on the regime where $U/t_h \ll 1$ at $T=0$.~The ground state is then described by the superfluid density order parameter and the quasi-particles are coherent long wavelength fluctuations in the phase.~A Gutzwiller ansatz in terms of on-site number states provides an adequate formulation of the wavefunction.~If we consider increasing the particle interaction $U$ towards the critical value $U_c$ that separates the superfluid from the Mott-insulating phase, we would expect to see an increase in depletion of the condensate.~The part of the wavefunction that describes particles outside the condensate has complete \emph{phase decoherence} so that the phase fluctuations at the $i^{th}$ site, for $N_i^\prime$ particles outside the condensate, can be described using the eigenstates of the phase operator in the number state basis~\cite{Popov1992}:
\begin{eqnarray}
 \left| \theta_m \right>  =  \sum_{n=0}^{N_i^\prime -1} \frac{{(e^{i  \theta_{m}} \, \hat{b}^\dagger_i )}^n}{n!} \, \left| 0 \right>_i  \label{eqn:quasipart}\end{eqnarray}
where $\left| 0\right>_i$ is the ground state wavefunction for the $i^{th}$ site.~With this definition, the $i^{th}$ site phase operator is $\hat{\theta}_i \!\equiv \! \sum_{m=0}^{N_i^\prime-1} \, \theta_m \left| \theta_m \right>_i\left< \theta_m \right|_i$
where $\theta_m\! = \!\theta_0\! + \!\frac{2 \pi m }{N^\prime}$ and $\theta_0$ is a reference angle which must be averaged over when computing observables.~In general, the depletion number $N_i^\prime$ is a function of the total number of particles $N_i$ and the interaction $U$.~When $N_i^\prime$ is large, this dependence on $N$ and $U$ can be incorporated into a coefficient $f_i(N, U)$ that multiplies each term of the sum while maintaining the relative sizes $N_i << N_i^\prime << 1$:
\begin{eqnarray}
\left|\theta\right>_i = \sum_{n=0}^{\infty} \frac{{( f_i \, e^{i  \theta_0} \, \hat{b}^\dagger_i )}^n}{n!} \, \left| 0 \right>_i =  \hat{D}(\alpha) \, \left| 0 \right>_i \; .
\end{eqnarray}
where $\alpha_i = f_i e^{i \theta_0}$ and we have used the symmetric form for the phase coherent state $\hat{D}(\alpha) \equiv e^{\alpha_i \hat{b}^\dagger_i - \alpha_i^* \hat{b}_i}$.~The normalized quasi-particle operators can now be defined having the correct Bogoliubov limit:
\begin{eqnarray}
  \hat{a}_i^\dagger=\hat{b}_i^\dagger \,  \hat{D}^\dagger(\alpha) /\sqrt{1+ |\alpha|^2} \, , \;   \hat{a}_i=\hat{D}(\alpha)\,  \hat{b}_i/\sqrt{1+ |\alpha|^2}   \; .
 \end{eqnarray} 
 With this definition, the average number of bosons forming a quasi-particle is $\left< \hat{b}^\dagger \hat{b} \right> \!= \!1 \!+\! |\alpha|^2$, where $|\alpha|$ encodes the degree to which the bare quasi-particle operators $\hat{b}$ and $\hat{b}^\dagger$ are dressed or \emph{fused} to the background phase.~When working at the lattice scale it is clear that the reference angle $\theta_0$ must be single-valued under one full rotation so that the dependence on the polar angle is $e^{i ( \theta_0 + \theta)}$. However, when translating to the continuum limit, the two triangular sublattices of the honeycomb lattice become identified with the same spatial point so that a $2\!-\!1$ mapping of parameters is needed consistent with the double covering map $p: SU(2) \!\to \!SO(3)$.~The continuum quasi-particle statistics can then be determined by computing the Berry phase (holonomy) by adiabatically transporting a quasi-particle through a suitable closed path and correctly accounting for the SU(2) structure of the single-particle states.~The path must remain within a degenerate subspace of the Hamiltonian so that we may isolate the geometric phase from the dynamical phase. The holonomy is $\gamma \equiv \mathrm{exp} \oint_Cd \theta \left< \Psi_\alpha(\theta) \right|\frac{d}{d\theta} \left| \Psi_\alpha(\theta)\right>$ with $ \left| \Psi_\alpha(\theta)\right> =    \frac{ \hat{b}^\dagger \,  \hat{D}^\dagger(\alpha) }{\sqrt{1+ |\alpha|^2}} \left| 0 \right>$ and $\alpha = f \, e^{ i (\theta_0 + \theta) /2}$.  A straight forward calculation yields $\gamma = \mathrm{exp}( i \pi  |\alpha|^2)$ demonstrating fermion exchange statistics when $|\alpha|^2=1$. 
\vspace{-.4pc}

\section{Physical Parameters and Regimes} We list first the fundamental dimensionful parameters that we use.~They are as follows: the average particle density $n_0$, the chemical potential $\mu$, the lattice spacing $a$, the s-wave scattering length $a_s$, the mass of the constituent bosons $M$, and the lattice well depth $V_0$.~Several relevant composite quantities may be constructed from these. These are the effective speed of light $c_l=t_h a \sqrt{3}/2 \hbar$, the sound speed $c_s=\sqrt{U n_0/M}$, the interaction strength $U=4 \pi \hbar a_s/M$, the healing length $\xi =  t_h a \sqrt{3}/ 2 \hbar n_0 U$, and the hopping energy $t_h=\int \! d^2r \, w^* \hat{H}_0 w$, where $t_h$ depends on $a$ and $V_0$, respectively, through the overlap of Wannier functions and the lattice potential inside $\hat{H}_0$.~Two fundamentally important constraints regarding these quantities should be stated.~First, in order to avoid reaching the Landau velocity at the band edge and creating unwanted excitations we require $c_l < c_s$, where $c_s$ is the sound speed.~Thus we require $t_h  a \sqrt{3}/2\hbar < \sqrt{U n_0/M}$ or $(t_h  a \sqrt{3}/2\hbar)(4 \pi \hbar^2 a_s \,  \bar{n}/M)^{-1/2} < 1$.~For $^{87}$Rb with $t_h = \hbar  \times 10^3\, \mathrm{Hz} \, ,\,  a= 0.5 \times 10^{-7} \, \mathrm{m} \, , \, a_s =  5 \times 10^{-9} \, \mathrm{m}\,, \, \bar{n} = 2 \times 10^{12} \, \mathrm{cm}^{-3}$, we get $c_l/c_s \lesssim 0.17$.~Second, in order for our long-wavelength approximation to be correct, we require the NLDE healing length $\xi \equiv  t_h a \sqrt{3}/ 2 \hbar n_0 U \gg a$; using the same values for the physical parameters, we find $\xi \approx 9.43\, a$.

Next, we discuss the physical regimes for our theory. First, consider the extreme weakly interacting regime, i.e., $n_0 U/t_h \ll 1$.~Then excitations of the ground state obey bosonic statistics.~Furthermore, at length scales much larger than the healing length, $\xi k \ll 1$ where $k$ is a characteristic quasi-particle momentum, excitations are comprised of correlated particle-hole pairs that propagate with a dispersion given by $E\propto k^{1/2}$.~This is a \emph{Bose gas} of composite particles in the sense that excitations of opposite spin are paired up (albeit non-locally) to form bosons.~In contrast, for $ \xi k  \gg 1$, excitations are particle-like which corresponds to the case where spin eigenstates are excited independently. These states reflect the bipartite structure of the lattice (multi-component) but are local objects and so also reflect the bosonic nature of the fundamental constituent particles. They exhibit a mixture of fermionic \emph{and} bosonic properties having a Dirac-like dispersion $\propto k$ but with quantum operators that obey bosonic statistics.~This is a \emph{hybrid Dirac-Bose gas}.~On the other hand for moderate interactions, $n_0 U/t_h \ne 0$ and close to the critical point on the superfluid side, higher order processes become significant and quasi-particles become highly non-local objects. Here quasi-particles are heavily ``dressed" in the presence of the background condensate and so acquire a geometric phase resulting in anticommutation relations for their associated creation and annihilation operators (see previous section).~In this regime, the Bogoliubov approximation fails to provide a good model but we expect that the Dirac spinor structure will remain robust against many-body effects with quasi-particles undergoing a renormalization due to many-body interactions similar to the case of graphene~\cite{Jafari2009}.~This describes an interacting \emph{Dirac gas}. 
\vspace{-.4pc}

\section{Uniformly Moving Condensate} Now we return to the RLSE and solve them for the simplest case of a uniform background $\Psi({\bf r}) \equiv  \sqrt{n_0} \, e^{i {\bf q} \cdot {\bf r}} ( 1  , C_0 )^T$,  where $C_0 \in \mathbb{C}$ contains a relative phase, $n_0$ is the average particle density, and ${\bf q}$ is the condensate momentum measured with respect to the Dirac point. In order to obtain the coherence factors and quasi-particle dispersion, we must then solve a $4\times 4$ eigenvalue problem; the RLSE yield
\begin{equation}
E_{\bf k} =c_l'  \hbar   {\bf q} \cdot {\bf k}   \pm  \sqrt{ ( c_l \hbar  k)^2   +     n_0  U c_l \hbar  k} \; .\label{eqn:e_k}  \end{equation}
In keeping with the usual Bogoliubov notation found in the literature, we may write $E_{\bf k} =(c_l'/c_l)    \, {\bf q} \cdot \vec{\epsilon}_k^{\,0}   \pm  E_k^0$  where $\vec{\epsilon}_k^{\,0} \equiv c_l \hbar {\bf k}$ is the single quasi-particle energy for zero interaction and $E_k^0 = \sqrt{ ( \epsilon_k^0)^2   +     n_0  U  \epsilon_k^{\,0} }$ is the quasi-particle energy for a static background.~The associated coherence factors can then be written as $|u_{k ,A(B)}| =  ( E_k^0  + c_l \hbar  k )/\sqrt{4  E_k^0  c_l \hbar  k}$, $| v_{k,A(B)}| = | u_{k,A(B)} | (+\to -)$. The full interacting Hamiltonian is given by $\hat{H}_{\mathrm{RLSE}}  =  \frac{1}{4} U  n_{0}^2 A + c_l \hbar  q -   \sum_{k}^{\prime} (   2 \epsilon_k^0  + n_0  U   ) +  \sum_{k}^{\prime} E_k  \hat{c}_k^\dagger  \hat{c}_k$ , where $A$ is the area of the plane.~The first three terms are the mean-field and quantum corrections to the condensate energy and the last accounts for the number of quasi-particles present in the system.~The constant $c_l'$ is defined in terms of the overlap integral between Wannier states at neighboring lattice sites by $c_l' = \sqrt{3} a \tau /2 \hbar$, where ${\bf \tau}_{A, B} \equiv - \int \! d{\bf r} \,  w_A^* \nabla   w_B$ and $\tau = | {\bf \tau}_{A, B}|$~\footnote{Note that whereas for the effective speed of light we have $[ c_l] = m\cdot s^{-1}$, in contrast $[ c_l'] = m^2 \cdot s^{-1}$ since ${\bf \tau}_{A,B}$ is an integral over the gradient operator rather than the Laplacian.}.

The low energy behavior of the uniform condensate has a rich structure.~The ${\bf q}={\bf 0}$ case corresponds to a condensate with zero crystal momentum measured from the Dirac point but with momentum ${\bf K}$ relative to the lowest Bloch state of the crystal.~The idea of a condensate in motion relative to its background has been treated in both free-space as well as the case of a moving background lattice~\cite{Morsch2006, Yukalov2009}.~Physically, the lattice potential is moving relative to the stationary condensate (laboratory frame). Two-body collisions reduce the momentum of some particles relative to the lattice (slowing down) and increase the momentum of others (speeding up) corresponding to a finite depletion of the condensate.~In the laboratory frame, a two-particle collision appears as one particle gaining a component of momentum to the left and the other a component to the right.~This is consistent with the well known particle-hole symmetry of the Dirac Hamiltonian: negative energy states can be interpreted as positive energy states that propagate in the opposite direction.~In our theory these are quasi-particles with momentum ${\bf K} - {\bf k}$ (for the {\bf K}-Dirac point) relative to the lowest Bloch state. 

For ${\bf q}\!\!=\!\!0$ then, we get $E_k^{(\pm)} \!\equiv\! \pm \! E_k^0 \!= \pm \sqrt{ ( \epsilon_k^0)^2  +  n_0  U  \epsilon_k^0 }$.~The two energy regimes evident here are separated by the condition $ c_l \hbar k / n_0 U  \equiv \xi k \approx 1 $. At short wavelength, $k \xi \gg 1 $ so that  $ E_k^{(\pm)} \approx \pm ( c_l \hbar  k + n_0 U/2)$, where the dominant first term reflects only the presence of the honeycomb lattice, while the second term is a small mean-field Hartree shift due to the interaction with the background.~When $k\xi \ll  1 $, we find $E_k^{(\pm)} \approx \pm  \sqrt{k/\xi}$. These are collective excitations induced by the particle interactions just above the condensate energy.~The presence of negative energy modes means that the condensate may lower its energy through spontaneous emission of radiation.~This process can be suppressed by introducing an anisotropy in the lattice by breaking the A-B sublattice degeneracy with a deeper optical lattice in one direction~\cite{zhuSL2007}.~This results in an additional term in the dispersion opening up a mass gap $2  m_\mathrm{eff}$ at the Dirac point.~For the negative energy modes we then have  $E_k^{(-)}(m_\mathrm{eff}) = 2m_\mathrm{eff} -  \sqrt{ ( \epsilon_k^0)^2   +     n_0  U  \epsilon_k^0 }$ so that excitations require a minimum momentum determined by $c_l  \hbar  k_\mathrm{min}= \sqrt{ 4 m_\mathrm{eff}^2 + n_0^2 U^2} $.~Alternatively, we can consider the effect of the confining potential: this sets a lower bound for quasi-particle energy given by $ |E_{k (\textrm{min})}^{(-)}| \sim    \sqrt{ ( c_l \hbar \, 2 \pi/R_\perp)^2   +     n_0  U \, c_l\hbar \,  2 \pi/R_\perp}$, where $R_\perp$ is the characteristic trap radius in the 2D plane. 
\vspace{-.4pc}

\section{Cherenkov Radiation} This usually refers to the anisotropic emission of electromagnetic radiation from a source whose speed exceeds the local speed of light in some medium~\cite{Jelley1958}. This concept generalizes to any source moving through a medium at a speed that exceeds the phase velocity of the elementary excitations of the medium.~For example, a BEC moving in the laboratory frame, or with respect to a background, will ``radiate" (emit particles) when its speed exceeds the sound speed. Moreover, the radiation will be emitted in a cone subtended by a specific angle in the direction opposite the motion of the BEC. The RLSE can be used to demonstrate this effect in the present context of a BEC in a honeycomb optical lattice. 

For a BEC with momentum ${\bf q}\!>\!{\bf 0}$ measured from the Dirac point, examination of the angular dependence of $E_k$ reveals an intriguing structure for the emission of Cherenkov radiation.  We observe the following properties for $E_k$.~(1) When $v < c_l$, where $v=c_l'  q$ is the condensate speed, all excitations have positive energy regardless of the angle of emission.~(2) When $v > c_l$, quasi-particle energies are positive only for emission angles (measured relative to ${\bf q}$) for which $\theta < \theta_c \equiv \text{cos}^{-1}(-c_l/v)$ while all other modes have negative energy corresponding to the emission of radiation in a backwards cone bounded by $ \theta_c$. When $v = c_l$, $\theta_c = \pi$ marks the onset of radiation, in which case radiation is only emitted in the direction opposite ${\bf q}$. This unique directional property of the radiation suggests an obvious detectable signature in the laboratory: a time-of-flight analysis of a BEC prepared with precise values of the parameters should show a predictable shift in the momentum distributions between the forward and backward directions. 
\begin{figure}
\vspace{0pc}
\begin{tabular}{c c}
\hspace{0pc}  \vspace{-1pc} \includegraphics[width=3.8cm]{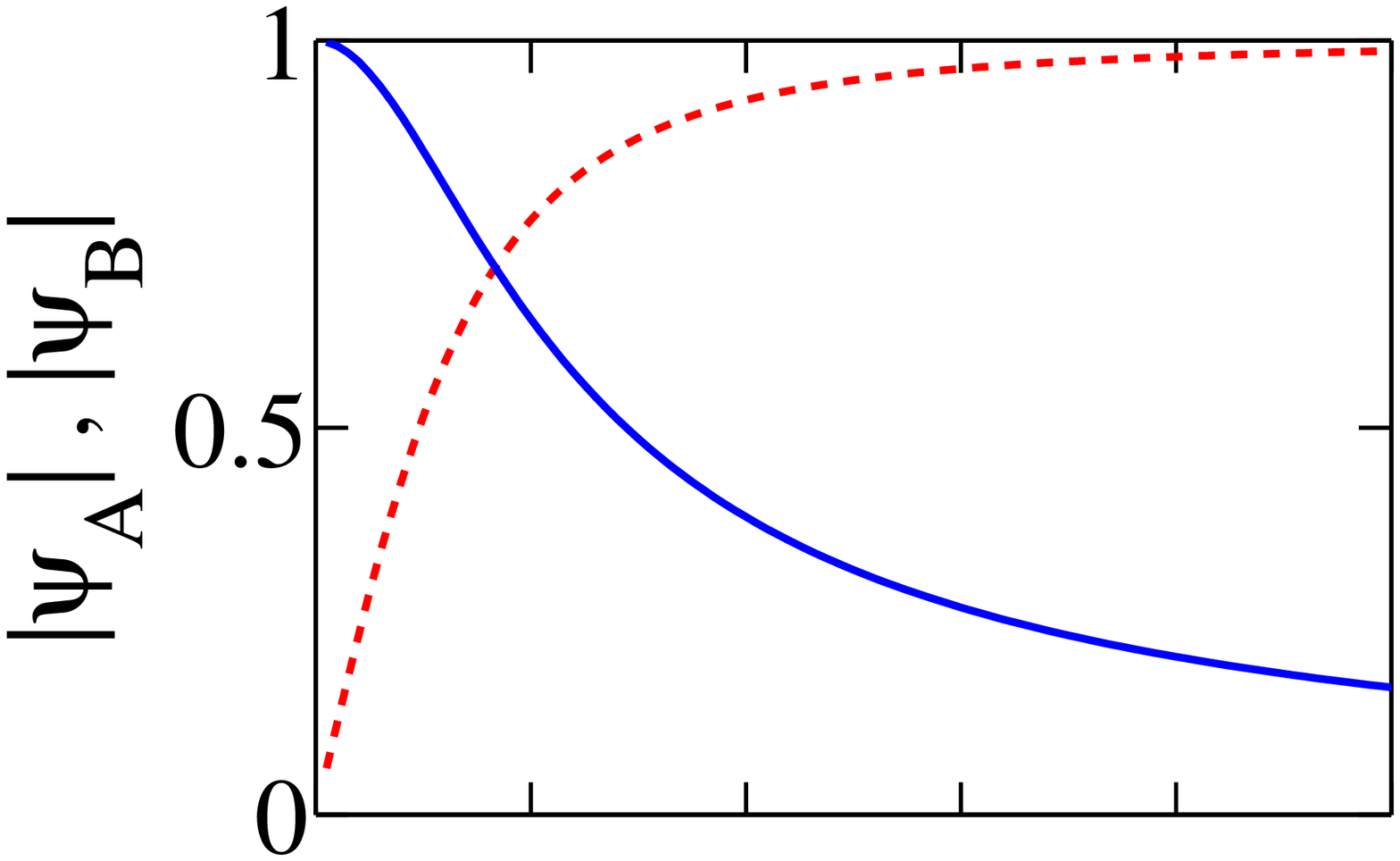} \hspace{0pc}&  \hspace{-1pc} \vspace{-1pc}  \includegraphics[width=3.8cm]{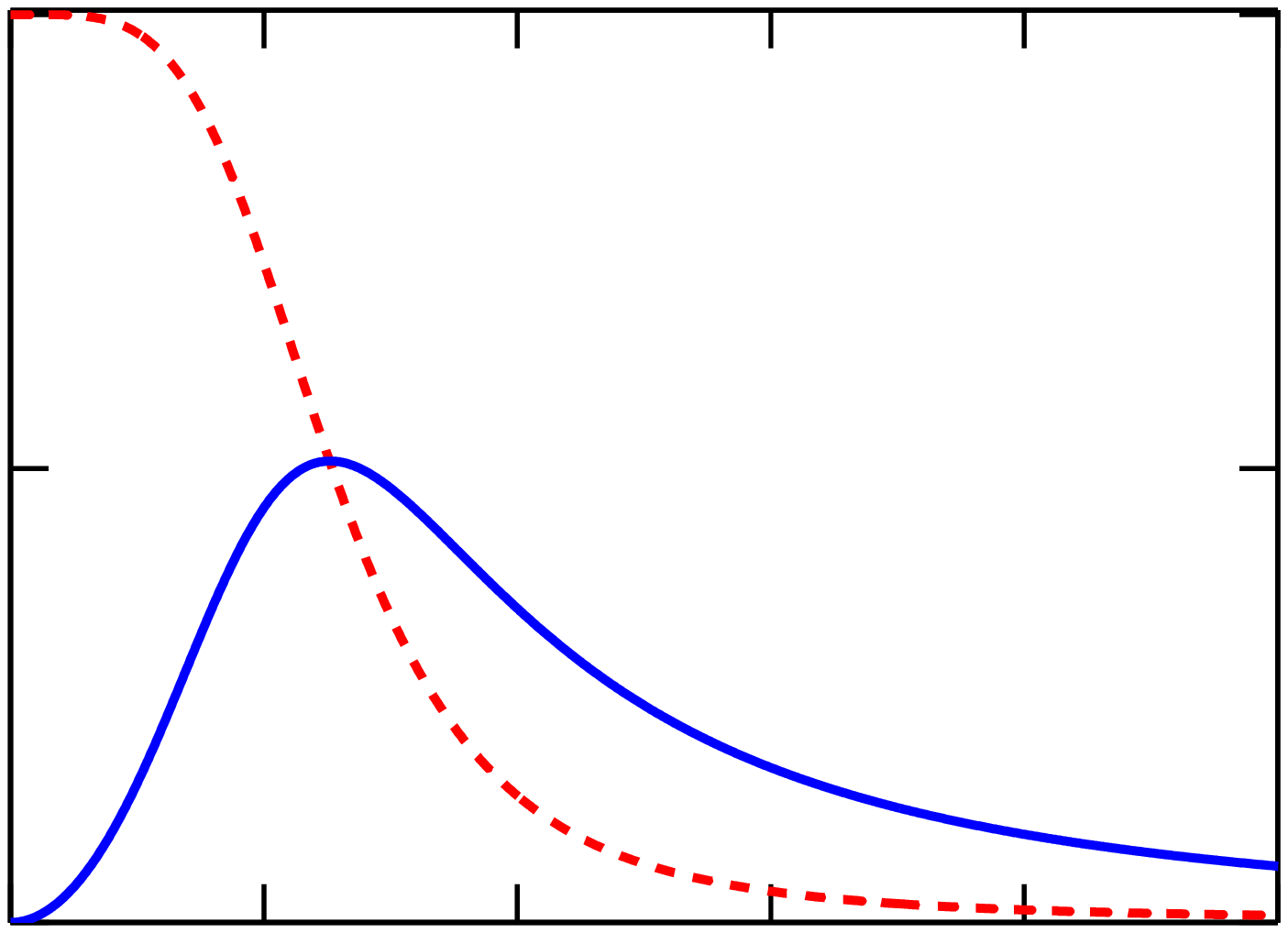} \vspace{0pc} \tabularnewline
 \vspace{0pc} \hspace{-8pc} (a) & \hspace{-9pc}  \vspace{0pc}(b)   \tabularnewline 
\hspace{0pc}  \vspace{-1pc} \includegraphics[width=3.8cm]{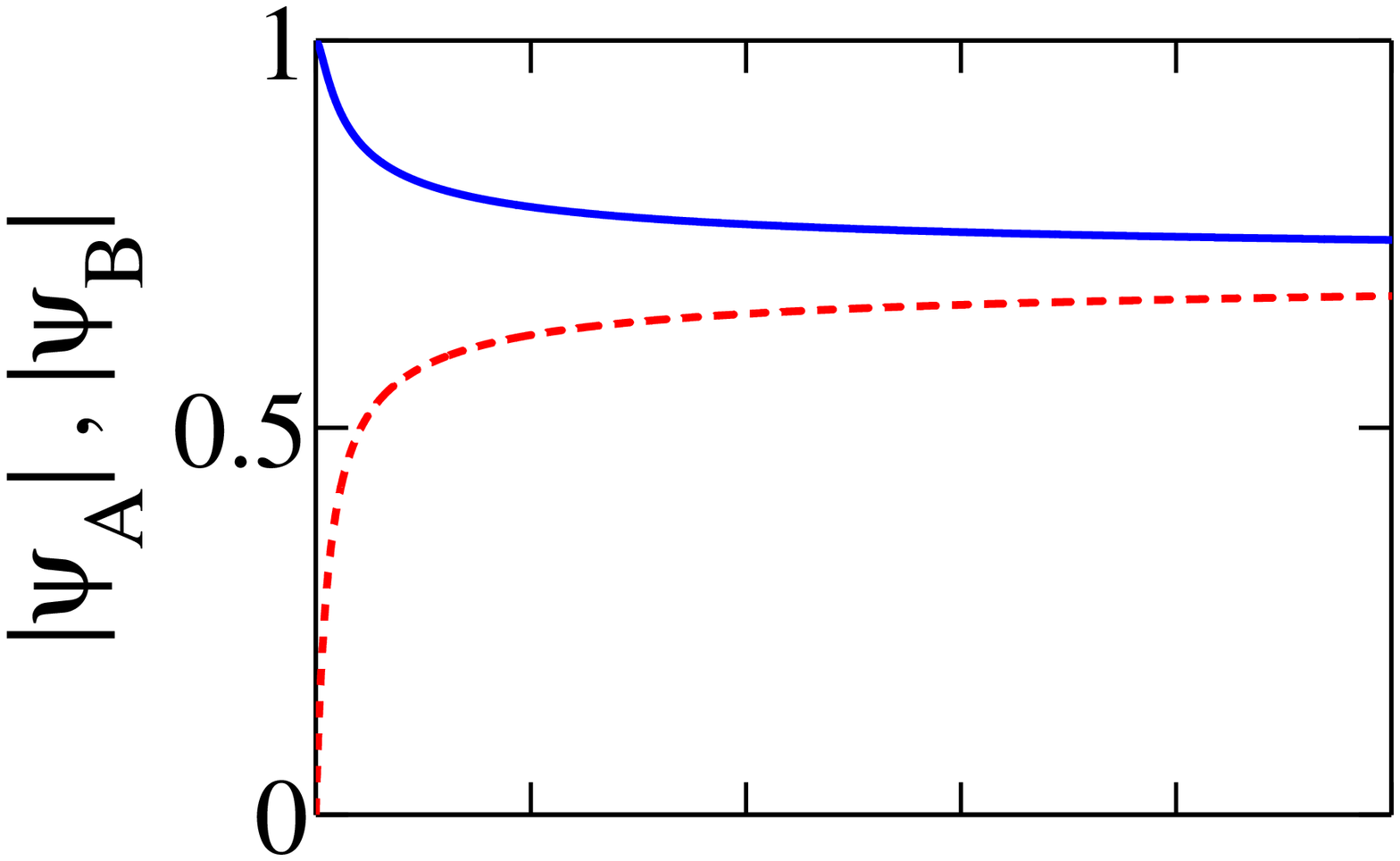} \hspace{0pc} &  \hspace{-1pc} \vspace{-1pc}   \includegraphics[width=3.8cm]{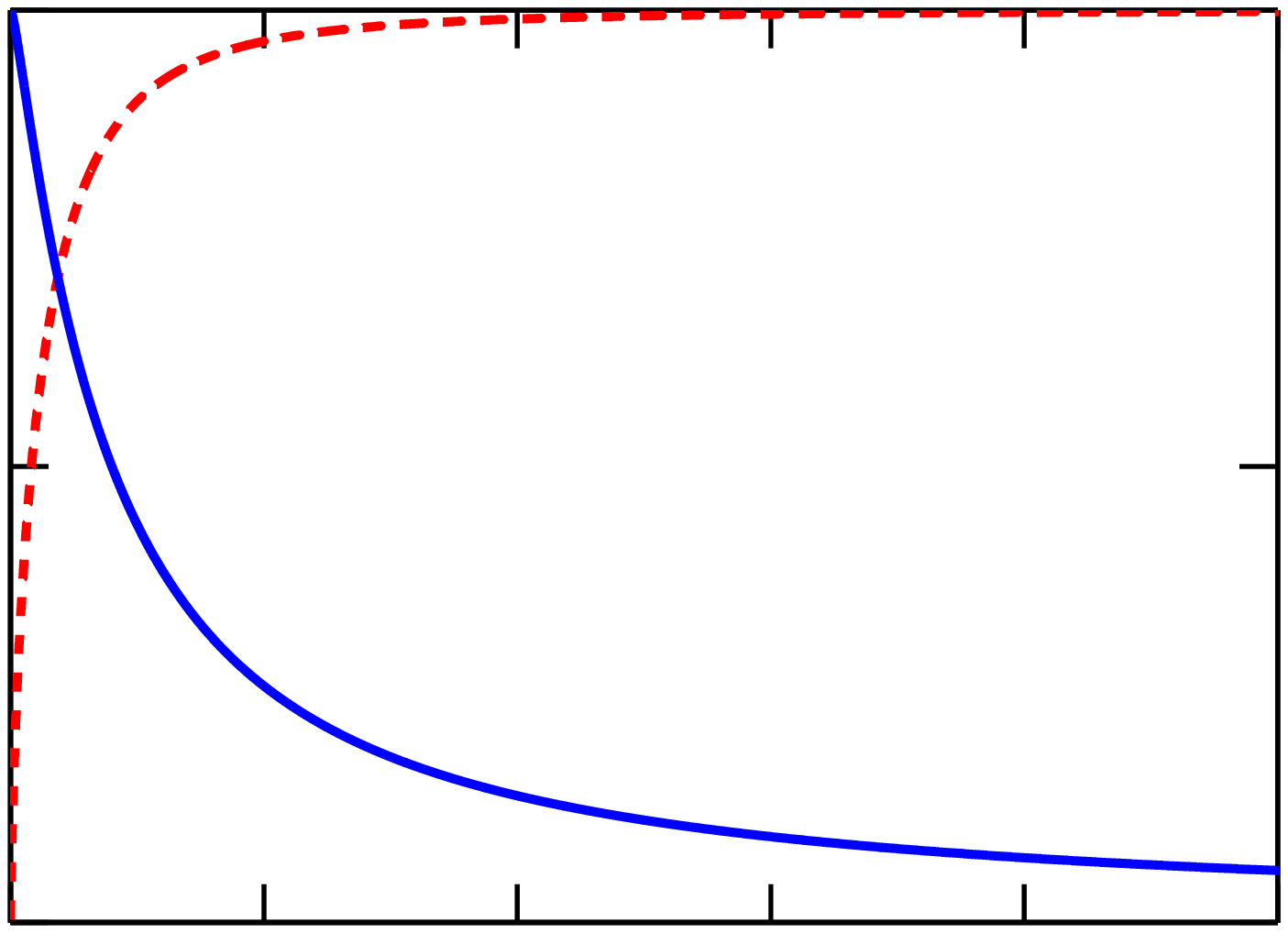}    \tabularnewline 
\hspace{-8pc} (c)  & \hspace{-9pc} (d)    \tabularnewline 
\hspace{0pc}  \vspace{-.9pc} \includegraphics[width=3.8cm]{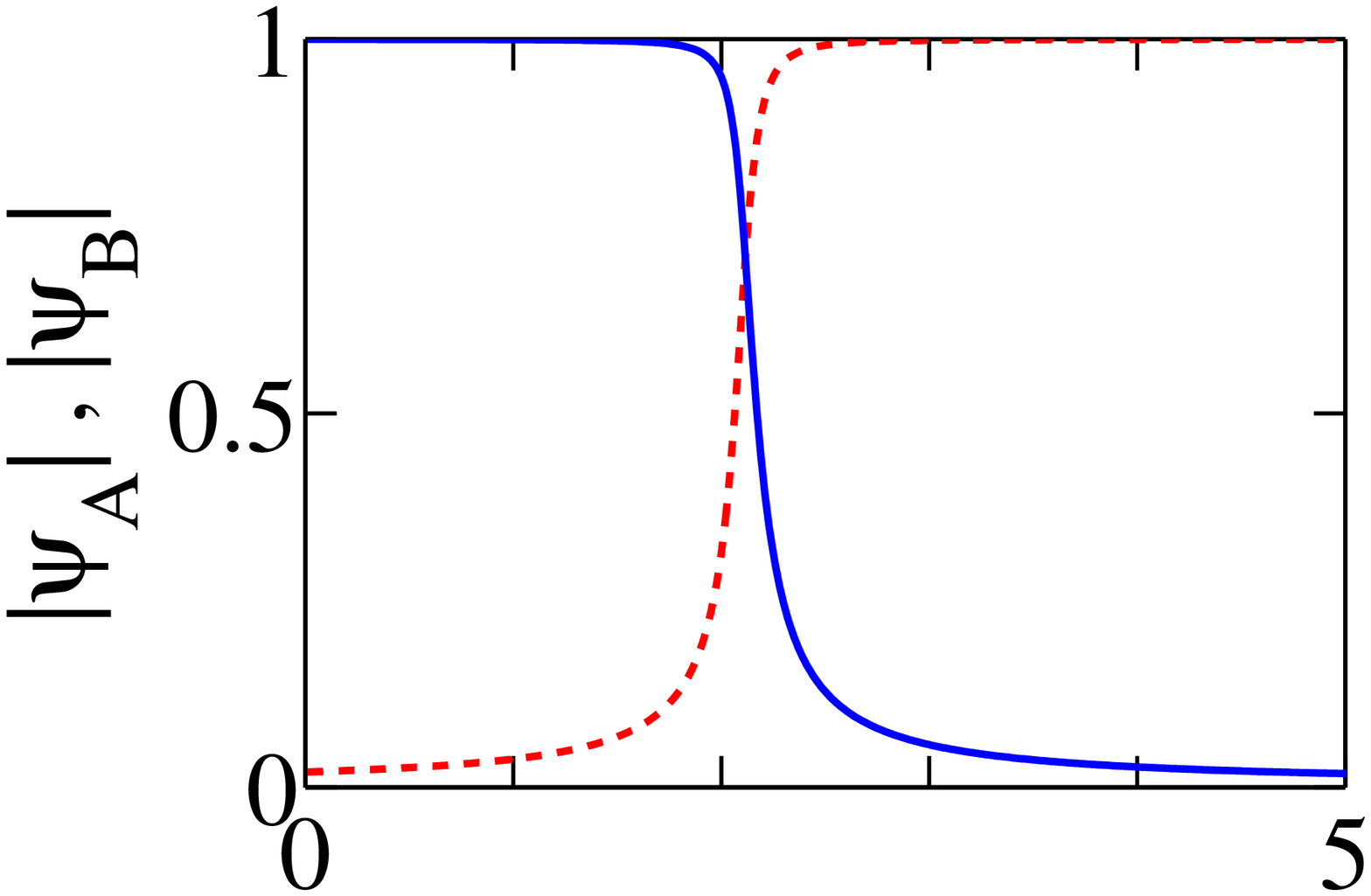} \hspace{0pc} &  \hspace{-1pc} \vspace{-.9pc}   \includegraphics[width=3.8cm]{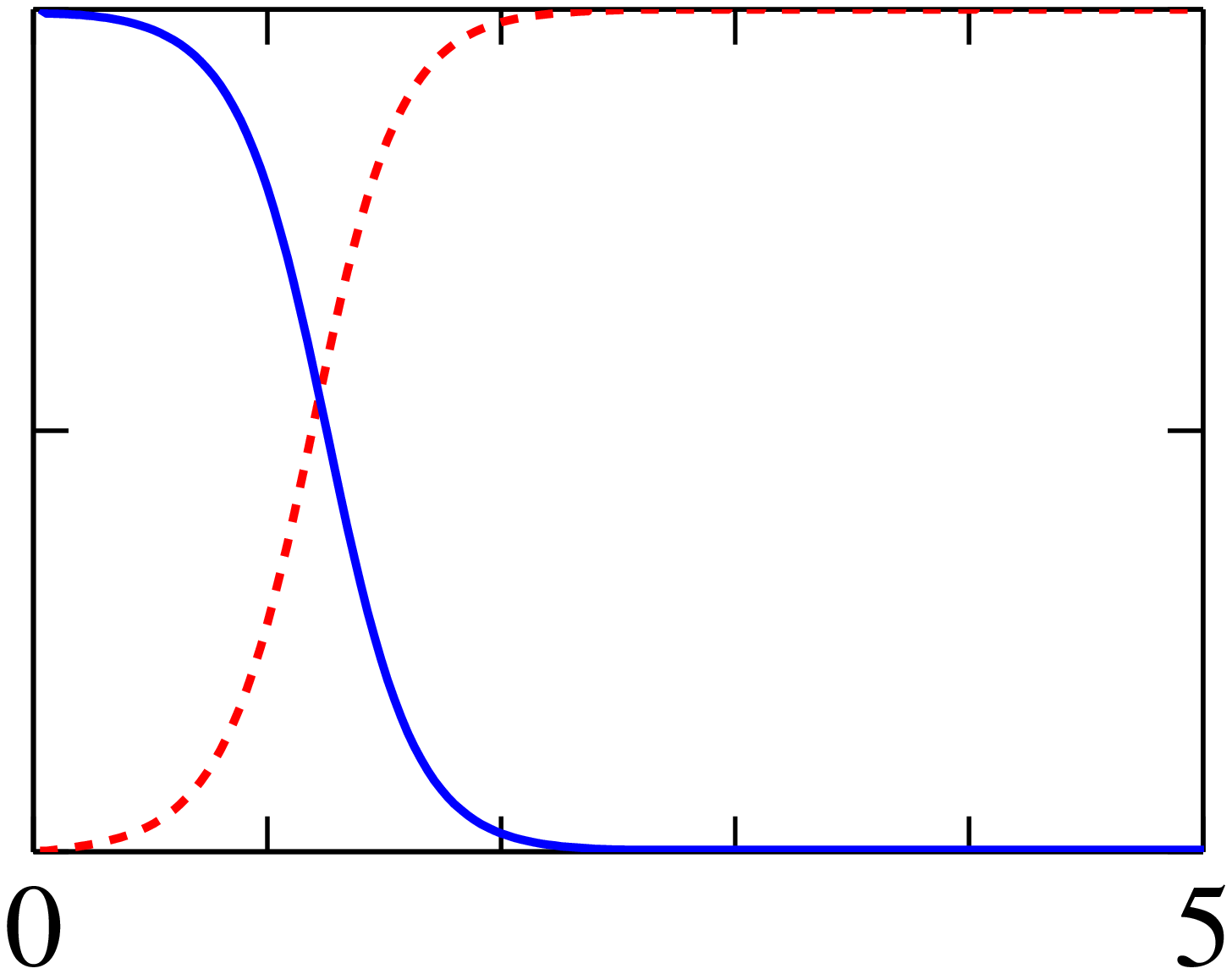}    \tabularnewline 
\hspace{-1.5pc} \vspace{-.001pc} (e) \hspace{5pc}$r\!/\!\xi$ & \hspace{-3.8pc} \vspace{-.001pc} (f)   \hspace{4pc}$ r \!/\!\xi$  \tabularnewline 
\end{tabular}\vspace{-.1pc}
\caption{\emph{Localized solutions of the NLDE},  $\Psi_A$~(red~dashed curves) and $\Psi_B$ (blue solid curves), in units of NLDE healing length $\xi$:~(a) vortex/soliton,~(b) ring-vortex/soliton,~(c) half-quantum vortex,~(d) planar skyrmion,~(e) line skyrmion, and~(f) line-soliton.} \label{fig.2}  \vspace{-1pc} 
\end{figure}

\section{Nonlinear Localized Modes} To obtain solutions of the NLDE that are localized in $x,y$ for $U\!>\!0$, we substitute the plane-polar ansatz $\Psi_A({\bf r})=c_A\,  \text{exp} [i p_A(\theta) ]  F_A(r)$, $\Psi_B({\bf r})=c_B \,  \text{exp} [ i p_B(\theta) ]  F_B(r)$ into the NLDE. Then $c_A=i$,  $c_B=1$, and there are two possible combinations for the angular functions: (i) $ p_A(\theta)= (l-1) \theta  , p_B(\theta)= l \theta \;; \; \text{(ii)} \; p_A(\theta)= (l-1/2) \theta  ,  p_B(\theta)= (l+1/2)\theta $, with $ l \in \mathbb{Z} $.~In particular, $ l=0$ in (i) corresponds to a vortex configuration in $\Psi_A$ filled in at the core with a nonzero soliton for $\Psi_B$.~Solutions of this type exist for different relative values of $\mu$ and $U$ and for several asymptotic values of the components: $\lim_{r\to \infty}(\Psi_A, \Psi_B) \in \{(-i\sqrt{\frac{\mu}{U}} , 0), ( -i\sqrt{\frac{\mu}{U}}, \sqrt{\frac{\mu}{U}} ),(0,0)\}$. For $l =1$, we obtain the same types of solutions but with $\Psi_A$ and $\Psi_B$ exchanged. For $l > 1$, centripetal terms are present for both $F_A(r)$ and $F_B( r)$ so that we must have $\Psi_A(0) = \Psi_B(0)= 0$ and both components are vortices with zero core densities. For the $l=1$ case, we also obtain a \emph{skyrmion} solution for which the pseudospin ${\bf S}=\bar{\Psi}({\bf r}) {\bf \sigma} \Psi({\bf r})$ (with Pauli vector ${\bf \sigma}$) exhibits an integral number of flips near the core and approaches a constant value far from the core. This feature is encoded in a topologically conserved charge $(1/8\pi)\int_\Omega \! d{\bf r}\, \epsilon^{ij} {\bf S} \cdot \partial_i {\bf S} \times \partial_j  {\bf S} $ which one recognizes as the Pontryagin index that classifies the mapping $S^1_{\mathrm{spin}} \to S^1_{\partial\Omega}$ where the two circles $S^1_{\mathrm{spin}}$ and $S^1_{\partial \Omega}$ parameterize the rotations between the densities $\rho_{A(B)}$ and the polar angle rotation on the 2D boundary $\partial \Omega$ at spatial infinity. In general, similar types of solutions exist for (ii) above. Analytical and numerical solutions are plotted in Fig.~\ref{fig.2} for which different values of $\mu/U$ and $l$ allow us to obtain the different asymptotic forms.

Besides vortices with integer phase winding, we also find solutions with fractional phase winding, called \emph{half-quantum vortices} (HQVs). Ordinarily, analyticity (single-valuedness) of the order parameter forbids the rotation of the phase of $\Psi$ around the core to take on fractional values. In the NLDE, $\Psi$ can acquire a coherent internal Berry phase in addition to an external phase whose angles are identified with the polar angle $\theta$~\cite{jiAnChun2008,lagoudakis2009}.~Such states may have half-integer winding in both the internal and external phase angles while remaining single-valued overall.~We obtain HQVs with asymptotic form $\lim_{r\to\infty}\Psi_{\mathrm{HQV}}({\bf r})\!\! = \!\!2 i  \sqrt{n_0/2} \, e^{-i \theta/2 }    [\cos(\theta/2)  ,  i  \sin(\theta/2) ]^\mathrm{T}$; the complete solution is shown in Fig.~\ref{fig.2}~(c).

We also obtain one-dimensional kink-soliton, skyrmion, and line-soliton solutions.~The kink and skyrmion solutions are obtained by a straight-forward substitution of the ansatz $\Psi(x) = \eta \, [\mathrm{cos}(\varphi)\, ,\, \mathrm{sin}(\varphi)]^T$ into the NLDE and then considering the distinct cases where $\varphi\!=\!\mathrm{constant}$ (kink) or $\eta\!= \! \mathrm{constant}$ (skyrmion).~The line-soliton solution is obtained when both $\eta$ and $\varphi$ are functions of $x$ with the additional condition that, at the origin, $\eta$ remains below a certain value.~This ensures that $\mu^2 <  \sqrt{U/8} < (U+1)/2U \, \Rightarrow \, U <   3.365 \, \mathrm{and}\, \mu^2 <  0.649$, which allows the wavefunction to collapse away from the $y$-axis while the nonzero wavefunction near and along the $y$-axis has a Lorentzian form in the $x$-direction due to the attractive effect from the kinetic terms. 
\begin{figure}
\begin{tabular}{c c}
 \hspace{-1.2pc} \includegraphics[width=4.268cm]{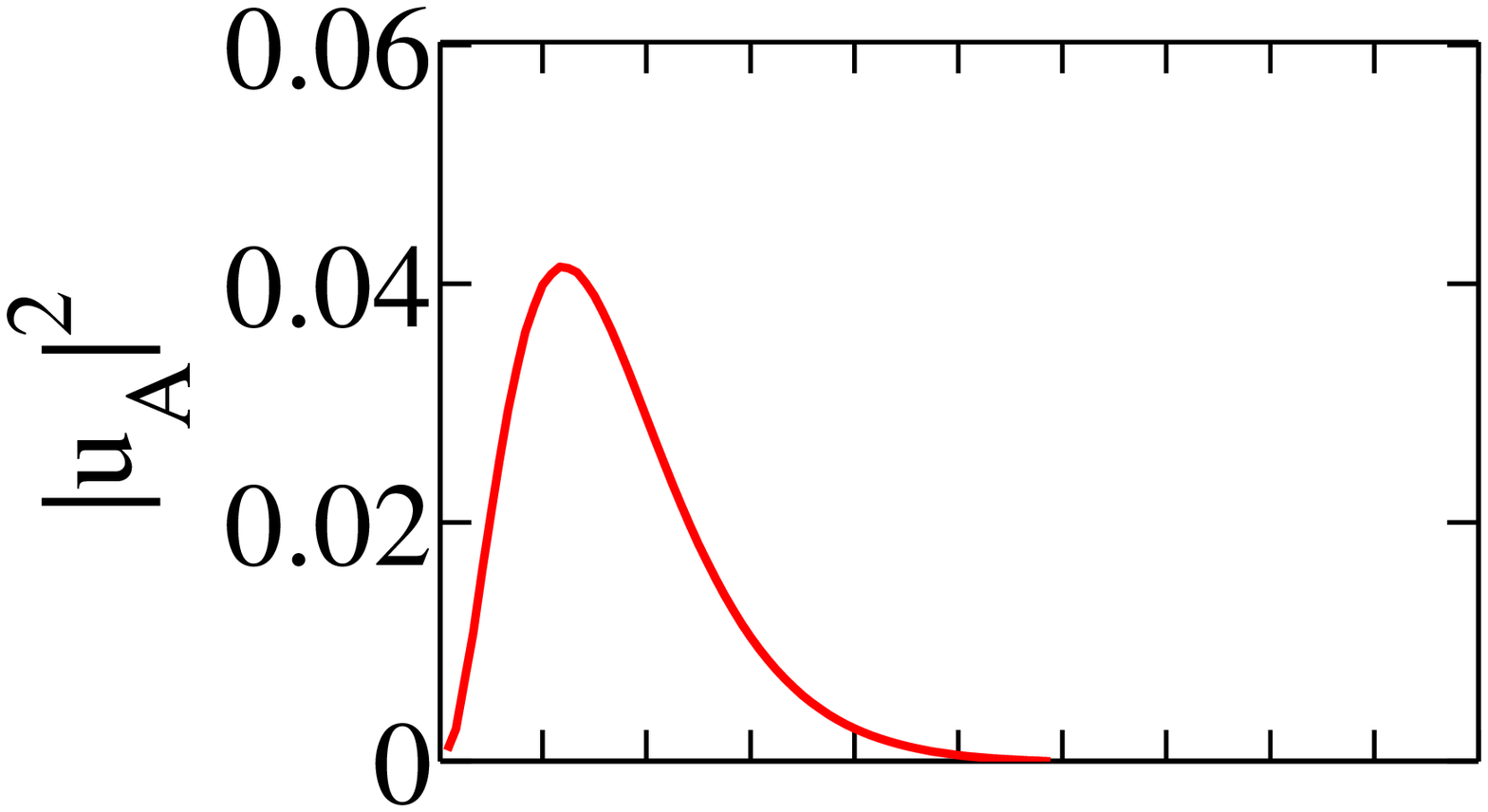} \hspace{-1pc} \vspace{0pc} &  \hspace{-.7pc}  \includegraphics[width=4.268cm]{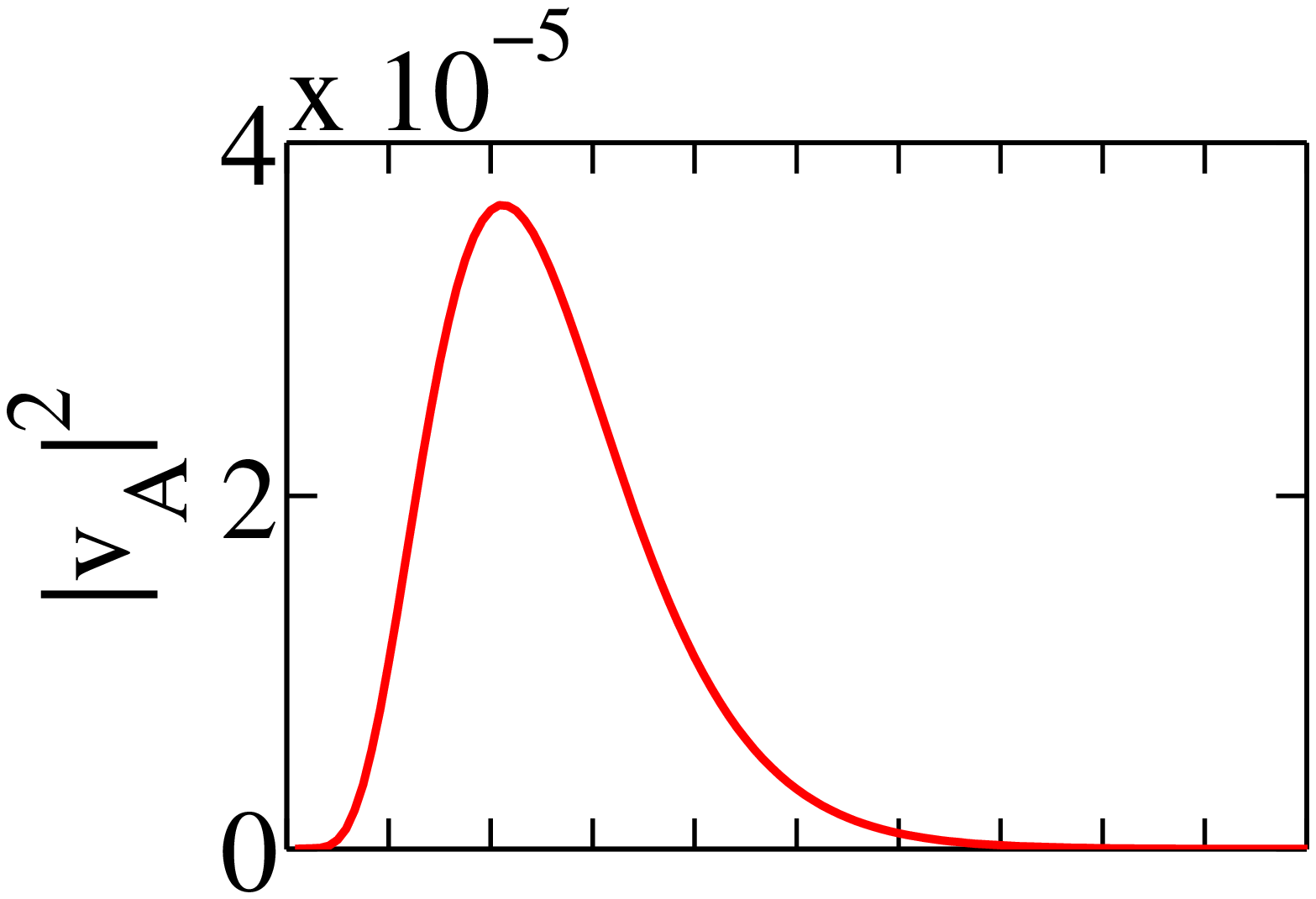} \vspace{-1.9pc} \tabularnewline
 \hspace{-8pc}  \vspace{-.08pc}  (a) & \hspace{-9pc}  \vspace{-.08pc}(b)   \tabularnewline 
\hspace{-1.8pc}  \vspace{-.9pc} \includegraphics[width=4.268cm]{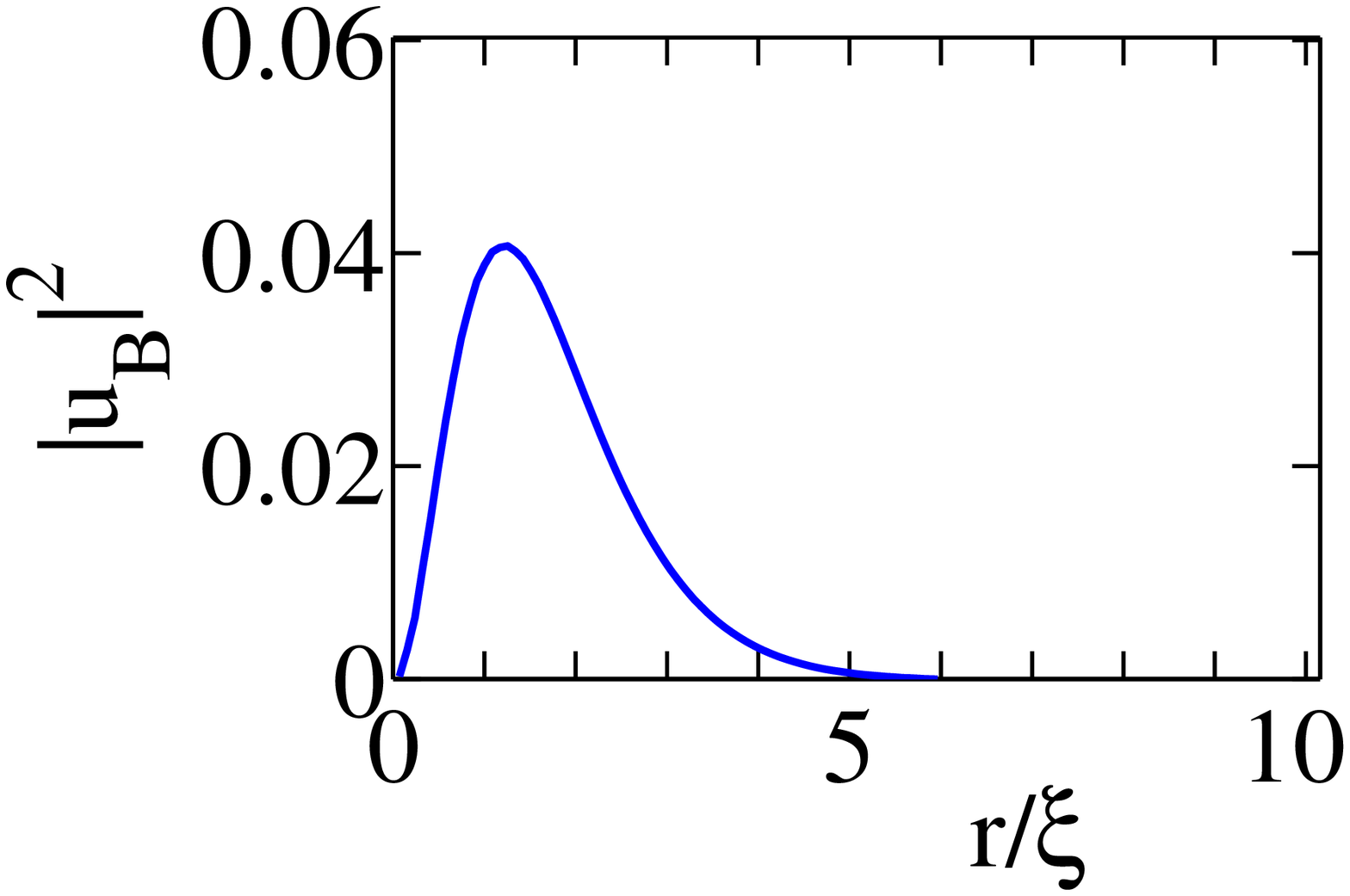} \hspace{-1pc} &  \hspace{-1pc} \vspace{-1pc}   \includegraphics[width=4.268cm]{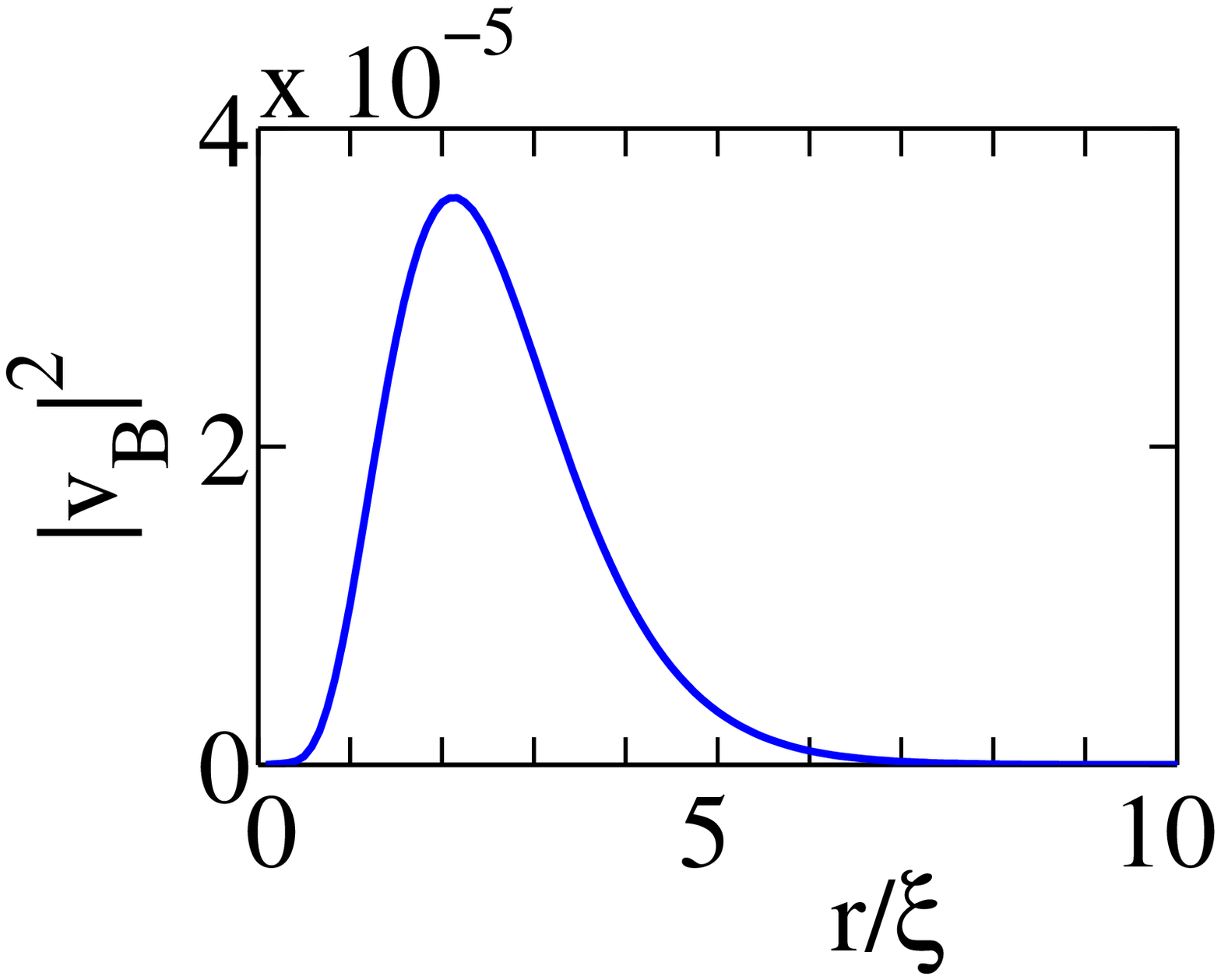}    \tabularnewline 
\hspace{-8pc} (c)  & \hspace{-9pc} (d)    \tabularnewline 
\end{tabular}
\caption{\emph{Plots of lowest quasi-particle excitation for the vortex/soliton configuration.} \label{fig.3}}
\vspace{-1.2pc}
\end{figure}
\vspace{-.4pc}

\section{Localized Mode Stability} To check the stability of our localized solutions we  substitute the corresponding solution into the RLSE and obtain the low energy spectrum.~For the case of the vortex/soliton ($ l = 0$), we solve the eigenvalue problem numerically and obtain $E_0\!=\!-3.9274+0.0020\, i$ in units of $n_0 U$.~Similar to the case of a trapped rotating BEC \cite{fetter2009}, we find that the lowest mode in the spectrum is \emph{anomalous} with negative energy and positive norm.~The coherence factors shown in Fig.~\ref{fig.3},
near the core of the vortex, appear as $|u_{k ,A(B)}|^2 \sim 10^{-2}$,  $| v_{k,A(B)} |^2 \!\sim \!10^{-5}$ so that $u_{k ,A(B)}>> v_{k ,A(B)}$.~This mode is dynamically unstable due to the presence of a small imaginary component which gives a decay rate relative to characteristic oscillation time of $0.0020/3.9274 = 0.0005$.~The lowest quasi-particle energies for the other localized solutions are: $- 3.9276 +  0.0019\, i $; $2.634 \times 10^2 + 9.96 \times 10^4i$; $- 3.9274 + 0.0019\, i$; $7.8409 \times 10^{-3}\! - 9.9993 \times 10^2\, i$; $7.9349 \times 10^{-3} \!- 9.9993 \times 10^2\, i$; for the ring-vortex/soliton, half-quantum vortex, planar skyrmion, line-skyrmion, and line-soliton respectively, where all quantities are given in units of $n_0U$.

In conclusion, we have shown that an effective relativistic fermionic system may be designed using ordinary cold bosonic atoms as the underlying degrees of freedom.~We solved the resulting NLDE for different classes of nonlinear modes including half-quantum vortices.~We derived and solved relativistic linear stability equations and gave explicit criteria for experimental observation of Cherenkov radiation, as well as predicting an anomalous mode for the vortex/soliton solution.~Density profiles may be observed by time-of-flight techniques to detect both massive and massless Dirac fermions in the laboratory~\cite{zhuSL2007,wuCongjun2008}; nonlinear modes involving phase winding can be created by techniques analogous to those used at JILA~\cite{matthews1999}; and we anticipate that Bragg scattering can be used to populate the Dirac cones at both ${\bf K}$ and ${\bf K}'$ points, leading to arbitrary superpositions over our localized solution types between the two cones, and thereby populating all four components of the Dirac spinor.

We thank Ken O'Hara, Jeff Steinhauer, and Michael Wall for useful discussions.~This work was supported by the NSF and the Aspen Center for Physics.


\begin{thebibliography}{0}

\bibitem{garay2000}
L.~J. Garay, J.~R. Anglin, J.~I. Cirac, and P. Zoller, Phys. Rev. Lett. {\bf
  85},  4643  (2000).

\bibitem{rapp2007}
A. Rapp, G. Zar\'and, C. Honerkamp, and W. Hofstetter, Phys. Rev. Lett. {\bf
  98},  160405  (2007).

\bibitem{yuYue2008}
Y. Yu and K. Yang, Phys. Rev. Lett. {\bf 100},  090404  (2008).

\bibitem{hartnoll2010}
S.~A. Hartnoll,  in {\em Understanding Quantum Phase Transitions}, edited by
  L.~D. Carr (Taylor \& Francis, Boca Raton, FL, 2010), Chap.~28.

\bibitem{snoek2009}
M. Snoek, S. Vandoren, and H.~T.~C. Stoof, Phys. Rev. A {\bf 74},  033607
  (2006).

\bibitem{merkl2009}
M. Merkl {\it et~al.}, Phys. Rev. Lett. {\bf 104},  073603  (2010).

\bibitem{carr2009g}
L.~H. Haddad and L.~D. Carr, Physica D: Nonlinear Phenomena {\bf 238},  1413
  (2009).

\bibitem{kevrekidisPG2008}
{\em Emergent Nonlinear Phenomena in {B}ose-{E}instein Condensates}, edited by
  P.~G. Kevrekidis, D.~J. Frantzeskakis, and R. Carretero-Gonz\'alez
  (Springer-Verlag, Berlin, 2008).


\bibitem{fetterAL1972}
A.~L. Fetter, Ann. Phys. {\bf 70},  67  (1972).


\bibitem{soltan-panahi2010}
P. {Soltan-Panahi} {\it et~al.}, e-print arXiv005.1276  (2010).


\bibitem{kusk2010}
J.~K. Block and N. Nygaard, Phys. Rev. A {\bf 81},  053421  (2010).

\bibitem{Lim2009}
L.-K. Lim, A. Lazarides, A. Hemmerich, and C. Morais Smith, Europhys. Lett. {\bf 88}, 36001 (2009).


\bibitem{feder2000}
D.~L. Feder {\it et~al.}, Phys. Rev. A {\bf 62},  053606  (2000).

\bibitem{muellerEJ2004}
E.~J. Mueller, Phys. Rev. A {\bf 69},  033606  (2004).

\bibitem{kasamatsu2005}
K. Kasamatsu, M. Tsubota, and M. Ueda, Phys. Rev. A {\bf 71},  043611  (2005).


\bibitem{pitaevskii1961}
L.~P. Pitaevskii, Sov. Phys. JETP {\bf 13},  451  (1961).


\bibitem{Popov1992}
V.~N. Popov, V.~S. Yarunin, J. Mod. Optics {\bf 39}, 1525 (1992).


\bibitem{Jafari2009}
S.~A. Jafari, The European Physical Journal B {\bf 68}, 537 (2009).


\bibitem{Morsch2006}
O. Morsch and M. Oberthaler, Rev. Mod. Phys. {\bf 78}, 179 (2006).

\bibitem{Yukalov2009}
V.~I. Yukalov, Laser Physics {\bf 19}, 1 (2009).



\bibitem{zhuSL2007}
S.-L. Zhu, B. Wang, and L.-M. Duan, Phys. Rev. Lett. {\bf 98},  260402  (2007).



\bibitem{Jelley1958}
{\em Cerenkov Radiation and its Applications},  J.~V. Jelley  
(Pergamon Press, 1958).


\bibitem{jiAnChun2008}
A.-C. Ji, W.~M. Liu, J.~L. Song, and F. Zhou, Phys. Rev. Lett. {\bf 101},
  010402  (2008).


\bibitem{lagoudakis2009}
K.~G. Lagoudakis {\it et~al.}, Science {\bf 326},  974  (2009).


\bibitem{fetter2009}
A.~L. Fetter, Rev. Mod. Phys. {\bf 81},  647  (2009).


\bibitem{wuCongjun2008}
C. Wu and S. Das~Sarma, Phys. Rev. B {\bf 77},  235107  (2008).


\bibitem{matthews1999}
M.~R. Matthews {\it et~al.}, Phys. Rev. Lett. {\bf 83},  2498  (1999).

\end{thebibliography}
\end{document}